\documentclass[a4paper,oneside,english,oldversion]{aa}
\usepackage[T1]{fontenc}
\usepackage[utf8]{inputenc}
\usepackage{color}
\usepackage{array}
\usepackage{verbatim}
\usepackage{multirow}
\usepackage{amstext}
\usepackage{graphicx}
\usepackage{esint}
\usepackage[authoryear]{natbib}
\usepackage{subscript}

\makeatletter

\providecommand{\tabularnewline}{\\}
\newcommand{\lyxdot}{.}

\newcommand{\tsup}[1]{\textsuperscript{#1}}
\newcommand{\mbh}{M_{\bullet}}
\titlerunning{Precision astrometry of compact radio sources in M15}

\@ifundefined{showcaptionsetup}{}{%
 \PassOptionsToPackage{caption=false}{subfig}}
\usepackage{subfig}
\makeatother

\usepackage{babel}
\begin{document}

\title{Precision astrometry of pulsars and other compact radio sources in
the globular cluster M15}

\author{Franz Kirsten\inst{1}\fnmsep\inst{2}\fnmsep\thanks{franz@astro.uni-bonn.de, Member of the International Max Planck Research School (IMPRS) for Astronomy and Astrophysics at the Universities of Bonn and Cologne}\and
Wouter Vlemmings\inst{3}\and Paulo Freire\inst{1}\and Michael Kramer\inst{1}\and
Helge Rottmann\inst{1}\and Robert M. Campbell\inst{4}}

\institute{Max Planck Institut für Radioastronomie (MPIfR), Auf dem Hügel 69,
D-53121 Bonn, Germany\and Argelander Institut für Astronomie (AIfA),
University of Bonn, Auf dem Hügel 71, D-53121 Bonn, Germany\and Department
of Earth and Space Sciences, Chalmers University of Technology, Onsala
Space Observatory, SE-439 92 Onsala, Sweden\and Joint Institute for
VLBI in Europe, Oude Hoogeveensedijk 4, 7991 PD Dwingeloo, The Netherlands}

\abstract{The globular cluster (GC) M15 (NGC 7078) is host to at least eight
pulsars and two low mass X-ray binaries (LMXBs) one of which is also
visible in the radio regime. Here we present the results of a multi-epoch
global very long baseline interferometry (VLBI) campaign aiming at
i) measuring the proper motion of the known compact radio sources,
ii) finding and classifying thus far undetected compact radio sources
in the GC, and iii) detecting a signature of the putative intermediate
mass black hole (IMBH) proposed to reside at the core of M15. We measure
the sky motion in right ascension ($\mu_{\alpha}$) and declination
($\mu_{\delta}$) of the pulsars M15A and M15C and of the LMXB AC211
to be $(\mu_{\alpha},\,\mu_{\delta})_{\text{M15A}}=(-0.54\pm0.14,\,-4.33\pm0.25)\,$mas$\,$yr$^{-1}$,
$(\mu_{\alpha},\,\mu_{\delta})_{\text{M15C}}=(-0.75\pm0.09,\,-3.52\pm0.13)\,$mas$\,$yr$^{-1}$,
and $(\mu_{\alpha},\,\mu_{\delta})_{\text{AC211}}=(-0.46\pm0.08,\,-4.31\pm0.20)\,$mas$\,$yr$^{-1}$,
respectively. Based on these measurements we estimate the global proper
motion of M15 to be $(\mu_{\alpha},\,\mu_{\delta})=(-0.58\pm0.18,\,-4.05\pm0.34)\,$mas$\,$yr$^{-1}$.
We detect two previously known but unclassified compact sources within
our field of view. Our observations indicate that one them is of extragalactic
origin while the other one is a foreground source, quite likely an
LMXB. The double neutron star system M15C became fainter during the
observations, disappeared for one year and is now observable again---an
effect possibly caused by geodetic precession. The LMXB AC211 shows
a double lobed structure in one of the observations indicative of
an outburst during this campaign. With the inclusion of the last two
of a total of seven observations we confirm the upper mass limit for
a putative IMBH to be M$_{\bullet}<500$ M$_{\odot}$.}


\maketitle

\section{Introduction }

\begin{table*}[t]
\caption{\label{tab:obs-details}Details of the observations}

\centering{}%
\begin{tabular}{ccccccc}
\noalign{\vskip-0.3cm}
\hline\hline &  &  &  &  &  & \tabularnewline
\noalign{\vskip-0.2cm}
 &  &  &  & \multicolumn{2}{c}{rms {[}$\mu$Jy$\,$beam$^{-1}${]}} & beam size \tabularnewline
Epoch & Date & MJD & Array & dirty{*} & cleaned & {[}mas x mas{]}\tabularnewline
\hline 
\noalign{\vskip\doublerulesep}
1 & 10 Nov 2009  & 55146 & JbWbEfOnMcTrNtArGb  & 5.1 & 3.1 & 3.3 x 6.4\tabularnewline
2 & 07 Mar 2010  & 55263 & JbWbEfOnMcTrNtGb  & 8.5 & 5.4 & 2.3 x 30.9\tabularnewline
3 & 05 Jun 2010  & 55352 & JbWbEfOnMcTrNtArGb  & 6.7 & 4.6 & 3.0 x 6.6\tabularnewline
4 & 02 Nov 2010  & 55503 & JbWbEfOnMcTrGb  & 11.2 & 7.4 & 2.1 x 26.2\tabularnewline
5 & 27 Feb 2011  & 55620 & JbWbEfOnMcTrArGb  & 4.9 & 3.1 & 3.3 x 6.9\tabularnewline
6 & 11 Jun 2011  & 55723 & WbEfOnMcTrArGb  & 5.8 & 3.8 & 2.3 x 6.2\tabularnewline
7 & 05 Nov 2011  & 55871 & JbWbEfOnMcTrArGb  & 5.2 & 3.3 & 3.1 x 7.0\tabularnewline
\hline\hline &  &  &  &  &  & \tabularnewline
\noalign{\vskip-0.3cm}
\multicolumn{7}{l}{{\tiny {*} applying natural weights without any cleaning}}\tabularnewline
\end{tabular}
\end{table*}

Pulsars, typically searched for and detected with single dish radio
telescopes, are rapidly rotating, highly magnetized neutron stars
(NSs). Their spin axis and magnetic field axis -- along which relativistic
charged particles are accelerated emitting cyclotron radiation --
are misaligned giving rise to the pulsar phenomenon. Being very stable
rotators, pulsars are used as accurate clocks to measure their intrinsic
parameters such as rotation period $P$, spin down rate $\dot{P}$,
and position. The fastest pulsars, the so-called millisecond pulsars
(MSPs, $P<30\,$ms), are the most stable rotators allowing for very
accurate tests of theories of gravity \citep[e.g. ][]{antoniadis2013,freire12}.\textcolor{red}{{}
}Roughly one half of all MSPs has been found in globular clusters%
\footnote{For a compilation of all globular cluster pulsars see the webpage
by Paulo Freire: http://www.naic.edu/\textasciitilde{}pfreire/GCpsr.html %
} (GCs) where the frequency of stellar encounters is high, favoring
the evolution of normal pulsars to MSPs through angular momentum and
mass transfer in a binary system \citep[e.g. ][]{bhattacharya91}.\textcolor{red}{{}
}In total, about 6\% of the 2302 currently known pulsars (as listed
on the ATNF webpage%
\footnote{http://www.atnf.csiro.au/research/pulsar/psrcat/, accessed November
18 2013%
}, \citealp{Manchester05}) reside in GCs with Terzan 5 and 47 Tuc
leading the field with 34 and 23 confirmed pulsars, respectively,
all but one being MSPs. Despite their rotational stability, disentangling
all parameters of GC pulsars through pulsar timing is sometimes difficult
due to the presence of the gravitational field of the GC. In those
cases, model independent measurements of intrinsic pulsar parameters
such as parallax, $\pi$, and proper motion, $\mu$, can improve the
overall timing solution. The ideal way to measure $\pi$ and $\mu$
purely based on geometry is through radio interferometric observations.

Here we report about multi-epoch global very long baseline interferometry
(VLBI) observations of the core region ($\sim\,$4 arcmin) of the
GC M15 (NGC 7078). This GC is one of the oldest (13.2 Gyr, \citealt{mcnamara04})
and most metal poor ({[}Fe/H{]}$=-2.40$, \citealt{sneden97}) GCs
known to reside in the Galaxy. It is host to eight known pulsars (four
of them being MSPs), one of which is in a binary system with another
neutron star (PSR B2127+11C, \citealt{anderson90}, \citealt{anderson1993}).
Four of the other seven pulsars are located in close proximity to
the cluster core (within $<4.5\,$arcsec $=0.2\,$pc at the distance
$d=10.3\pm0.4\,$kpc, \citealt{vandenbosch06}) making them ideal
candidates to study cluster dynamics. In the same region, two low
mass X-ray binaries (LMXBs, thought of as progenitors to MSPs,\textcolor{red}{{}
}e.g. \citealp{tauris06} and references therein) have been reported
(\citealt{Giacconi74,Auriere84,white01}). One of them, 4U 2129+12
(AC211), is also detectable as a compact source in the radio regime.
This relatively high concentration of compact objects that have been
or currently are in a binary system is already indicative of the high
stellar density within the core region of M15. In fact, the observed
central brightness peak and the stellar velocity dispersion profile
gave rise to speculations that M15 could host an intermediate mass
black hole (IMBH, e.g. \citealt{newell76}). The predicted IMBH mass
$\mbh=1700^{+2700}_{-1700}$ \citep{gerssen03} has, hower, been ruled
out by \citet{kirsten2012}. Alternatively, a collection of $\sim1600$
dark remnants such as stellar mass black holes, NSs, and white dwarfs
in the central region of M15 could drive cluster dynamics (\citealt{baumgardt03,mcnamara03,murphy11}).
Based on the 1.5 GHz radio luminosity and assuming a minimum pulsar
luminosity of $2\,\mu$Jy, \citet{sun02} estimate that M15 could
host up to $\sim300$ pulsars beaming towards Earth.

In this project, we accurately measure the proper motion of all compact
objects detectable within our field of view and monitor their variability.
Apart from the eight pulsars and the LMXB AC211, two further compact
radio sources were reported previously by \citet{machin90} and \citet{knapp96}.
Those authors could, however, put no tight constraints on those sources'
(non-) association with the cluster. Furthermore, we look for previously
undetected compact objects within the observed region that might turn
out to be pulsars. 

The double neutron star system M15C has shown a number of unusual
glitches which need to be fitted with a number of parameters that
are highly covariant with fits for the proper motion. In particular,
the measurement of the orbital period decay caused by the emission
of gravitational waves is influenced by an acceleration in the cluster
potential and by a contribution due to a transverse motion (``Shklovskii
effect'', see \citealp{lorimer05}). Thus determining the transverse
motion of the pulsar will allow a better measurement of the line of
sight acceleration of the system within the cluster potential. Once
the proper motion is determined independently of any model, the covariances
in the fits to the timing model can be removed, improving the measurement
of all relativistic parameters.

M15A is very close to the core and has a negative period derivative,
which implies it is accelerating at a fast rate in the cluster's potential.
This acceleration rate has now been shown to vary with time \citep{jacoby06}.
The detailed variation is of great interest to investigate the gravitational
potential in the cluster center, but if we have only the timing it
must be disentangled from the proper motion signal. Therefore, an
independent estimate of the proper motion of the pulsar will allow
a much less ambiguous interpretation of the variation of the acceleration
of this pulsar.

In the following we will first describe the data taking and data reduction
process in section \ref{sec:Observations-and-data}. The data analysis
strategy and the results are the subject of section \ref{sec:Results}
while section \ref{sec:Discussion} deals with the discussion of the
implications of these results. The main findings of this project are
briefly summarized in section \ref{sec:Conclusions}.

\section{Observations and data reduction\label{sec:Observations-and-data}}

\begin{table*}
\caption{\label{tab:Correlation-Centers}Correlation Centers throughout M15}

\centering{}%
\begin{tabular}{ccc}
\noalign{\vskip-0.3cm}
\hline\hline &  & \tabularnewline
\noalign{\vskip-0.2cm}
Name & Ra & Dec\tabularnewline
\hline 
\noalign{\vskip\doublerulesep}
M15 (epoch 1 only) & 21:29:58.3500 & 12:10:01.500\tabularnewline
AC211 & 21:29:58.3120 & 12:10:02.679\tabularnewline
15C & 21:30:01.2034 & 12:10:38.160\tabularnewline
S1 & 21:29:51.9025 & 12:10:17.132\tabularnewline
VRTX1 & 21:29:56.3050 & 12:11:01.500\tabularnewline
VRTX2 & 21:29:56.3050 & 12:09:11.500\tabularnewline
VRTX3 & 21:30:02.4410 & 12:09:11.500\tabularnewline
\hline\hline &  & \tabularnewline
\end{tabular}
\end{table*}

\begin{figure}
\centering{}\includegraphics[width=0.9\columnwidth]{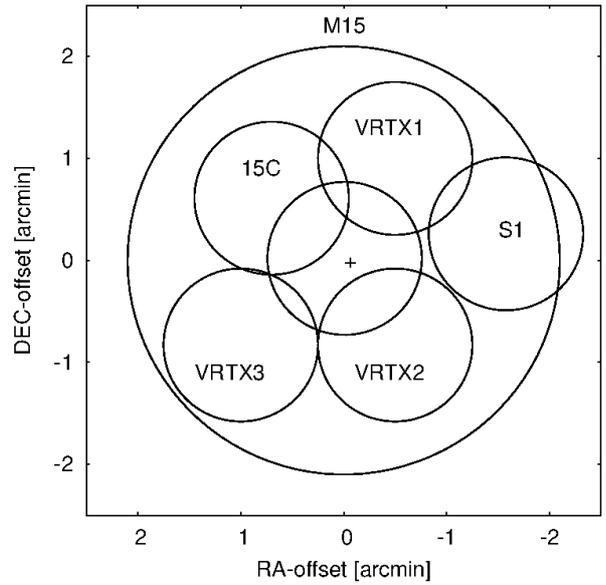}\caption{\label{fig:field of view}The primary beam of the entire array (large
circle, constrained by the Arecibo primary beam) and the position
and FOV of the six correlation centers relative to the pointing center
indicated by the cross at (0,0). The central circle with no label
corresponds to the correlation center labeled AC211 in Table \ref{tab:Correlation-Centers}.
The radius of the FOV is defined as to allow a maximal loss of 10\%
in the response of a point source caused by time and bandwidth smearing.}
\end{figure}

\begin{figure}
\begin{centering}
\includegraphics[width=0.85\columnwidth]{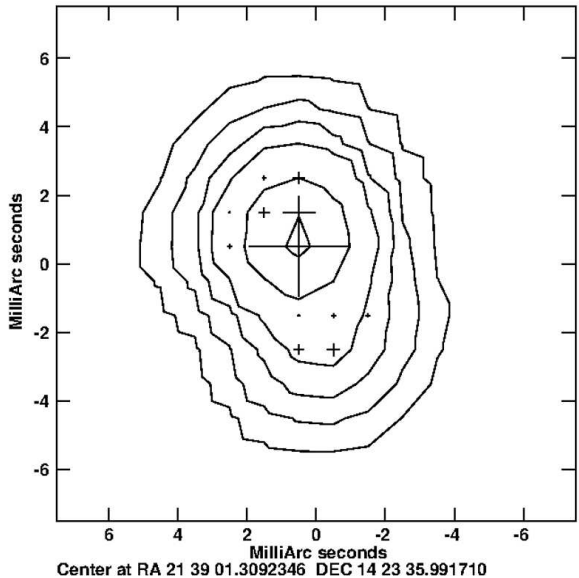}
\par\end{centering}

\caption{\label{fig:Model-of-the-phase-calibrator}Model of the phase calibrator
used in the final calibration. For better illustration the CLEAN components
used (marked by the crosses, the size of each cross indicates the
relative associated flux density) were convolved with a circular beam
of 2.6 mas FWHM (maximal resolution in right ascension). Contours
are (10, 50, 150, 300, 900, 1800) times the rms of 0.46 mJy. Note
the slightly extended structure in NE-SW direction.}
\end{figure}

\subsection{Observations\label{sub:Observations}}

We observed M15 seven times at 1.6 GHz with a global VLBI array that
included dishes of the European VLBI Network (EVN; Effelsberg, Jodrell
Bank, Onsala, Noto, Toruń, Westerbork, Medicina) plus the Arecibo
and the Green Bank Telescope (Table \ref{tab:obs-details}). The field
of view (FOV) of about 2 arcmin (constrained by the Arecibo primary
beam since only one dish of Westerbork was used) is large enough to
cover all of the know compact sources in the cluster. The observing
schedule encompassed six hours per epoch and was set up such that
the target cluster M15 and the phase reference source, the quasar
J2139+1423, located at $\sim3.17$ degrees towards the north-east
of the pointing center, were observed in an alternating fashion: After
each 3.5 minute scan of M15 we observed J2139+1423 for roughly 1.2
minutes. Altogether, the total integration time on M15 amounts to
about 3.6 hours in each of the seven epochs, roughly 1.5 hours of
which Arecibo was able to observe the GC. Due to technical failure
no Arecibo data is available for epochs two and four. 

All data \textbf{were} correlated at the Joint Institute for VLBI
in Europe (JIVE). Epochs one to five were correlated on the EVN-MkIV
correlator \citep{schilizzi01} while epochs six and seven were correlated
on the EVN software correlator at JIVE (SFXC%
\footnote{http://www.jive.nl/jivewiki/doku.php?id=sfxc%
}, Keimpema 2014, in preparation). In the first epoch, the correlation
center is the same as the pointing center at RA = 21\tsup{h}29\tsup{m}58\fs350,
Dec = 12\degr10\arcmin01\farcs500 (J2000). In order to avoid bandwidth
and time smearing the first epoch was correlated at a spectral resolution
of 512 channels per each of eight intermediate frequencies (IFs, bandwidth
of 16 MHz each, dual polarisation) and at a temporal resolution of
0.25 sec. The size of the final data set (230 GB), however, made further
data processing very slow. Consequently, subsequent epochs two to
seven were still observed at the same pointing center but correlated
at six different positions centered on or close to sources detected
in epoch 1 (Table \ref{tab:Correlation-Centers}). By using a much
lower spectral (128 channels per IF) and temporal (integration time
0.5 sec) resolution, the size of each of the six data sets amounts
to roughly 25 GB in each epoch. The FOV is $\sim0.75$ arcmin (<10\%
smearing) for the individual centers which is large enough to cover
most of the primary beam of Arecibo (Figure \ref{fig:field of view}).

\subsection{Data reduction and calibration\label{sub:Data-reduction-and-calibration}}

\begin{figure}
\begin{centering}
\includegraphics[width=0.85\columnwidth]{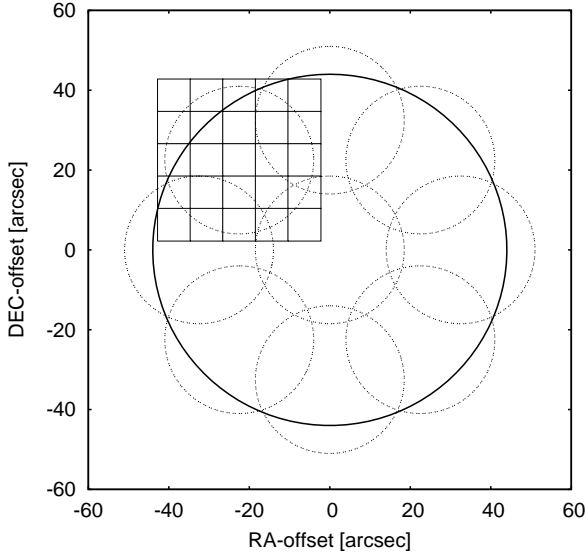}
\par\end{centering}

\caption{\label{fig:Imaging-strategy}Imaging strategy. The large solid circle
indicates the FOV of one sub-center, the small dashed circles represent
the relative position and the FOV of the shifted and averaged data
sets. The grid of rectangles illustrates the size and relative position
of each produced image. See text for details.}
\end{figure}

All data reduction and calibration steps are performed using the NRAO
Astronomical Image Processing System (AIPS). The EVN pipeline%
\footnote{http://www.evlbi.org/pipeline/user\_expts.html%
} provides the a priori system temperature and gain curve corrections
as well as flags concerning off-source times and band edges. We apply
these as given after which we correct the parallactic angle in CLCOR
and also compute ionospheric corrections in TECOR with the help of
total electron content maps as published by the Center for Orbit Determination
in Europe%
\footnote{ftp://ftp.unibe.ch/aiub/CODE/%
}.

In the next step we eliminate radio frequency interference (RFI).
Due to the data volume we do not flag manually but instead use the
AOFlagger \citep{offringa10,offringa12}. This software package works
with measurement sets which is why we first export the a priori calibrated
data sets (target source and calibrator sources individually) to a
fits file which is then read into the Common Astronomy Software Application
package (CASA) to produce measurement sets. After successful automatic
flagging we export each measurement set back to fits format in CASA
during which all flags are applied to the data. The flagged data sets
are then loaded back into AIPS. These steps are performed for each
calibrator source and all correlation centers individually in each
epoch. In cases where entire antennas, scans, or IFs were affected
by RFI our flagging-strategy was insufficient to account for it. Therefore,
we perform a further manual flagging step on the data averaged over
all channels in each IF.

In a first calibration step we compute bandpass solutions, phase alignments
and fringe rates independently for each epoch assuming a simple point
source model for the phase calibrator. In epochs \textcolor{black}{3
and 7 the data of the bandpass calibrator source, the quasar 3C454.3,
was affected by strong RFI resulting in a loss of considerable amounts
of data. Therefore, for consistency reasons, we compute bandpass calibration
tables from the phase calibrator source J2139+1423 using the entire
time range in the individual epochs. For phase alignment between bands,
we select a 30 sec time interval also from a scan of the phase calibrator.
After fringe fitting the data with solution intervals of 1.5 }min
we also self-calibrate on\textcolor{black}{{} J2139+1423}. In self-calibration
we solve for both amplitude and phase at solution intervals of 1.5
min. \textcolor{black}{In each epoch we detect the phase calibrator
at a signal to noise ratio (SNR) of }800--2200. 

\textcolor{black}{For the main calibration procedure, we produce a
global phase calibrator model concatenating the calibrated data of
J2139+1423 of all seven epochs. We use the clean components of the
image as the global model for further calibration (Figure \ref{fig:Model-of-the-phase-calibrator}).
We eventually rerun all of the calibration steps (except for self-calibration)
on the phase calibrator but instead of assuming a simple point source
model we apply our global model. }

The angular size of the primary beam allows us to detect the strong
unclassified source S1 \citep{machin90,johnston91}. This object lies
$\sim1.5$ arcmin to the west of the cluster core. We detect it at
peak flux densities of 2--4 mJy$\,$beam$^{-1}$; at this strength
it can serve ideally as in-beam calibrator eliminating residual phase
errors caused by atmospheric differences in the lines of sight to
phase calibrator and target cluster.\textcolor{black}{{} S1 is most
likely of extragalactic origin (see section \ref{sub:Extragalactic-origin-of-S1})
and, therefore, we select the image with the highest SNR (epoch 3)
as model for further in-beam self-calibration. Thus the position of
S1 is fixed to} RA = 21\tsup{h}29\tsup{m}51\fs9034555, Dec = 12\degr10\arcmin17\farcs13240.
\textcolor{black}{ Finally, we self-calibrate all epochs applying
this model.}

All calibration steps are performed on the data set named S1 in Table
\ref{tab:Correlation-Centers}. Once all solution (SN) tables are
computed they are combined into one final calibration (CL) table.
Since the pointing center is identical for all six correlation centers,
the calibration tables obtained for one sub-center can easily be applied
to the other ones speeding up data processing significantly. This
is true for all epochs except epoch 1. Due to the different observational
setup and correlation strategy, bandpass calibration, phase alignment
and fringe rate solutions are applied to the entire data. Before self-calibration
on S1 we shift the data to the six different correlation centers of
the subsequent epochs. This step is done using the technique described
in \citet{morgan11}. Afterwards, we average the six data sets to
the same spectral and temporal resolution as in epochs 2 to 7 and
continue with self-calibration on S1.

\subsection{Imaging\label{sub:Imaging}}

\begin{figure}
\begin{centering}
\includegraphics[width=0.8\columnwidth]{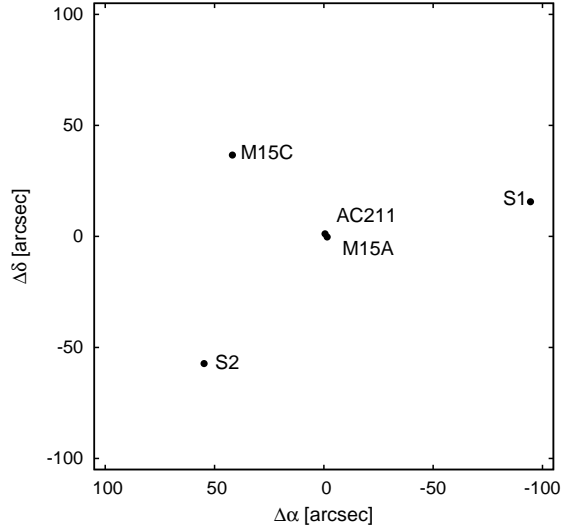}
\par\end{centering}

\caption{\label{fig:Distribution-of-sources}Distribution of the detected sources
relative to the pointing center.}
\end{figure}

\begin{table*}
\caption{\label{tab:Details-of-the-astrometric-fits}Details of the astrometric
fits for all detected sources. For the proper motion results and the
goodness-of-fit, $\chi_{\text{red}}^{2}$, the results for a fixed
parallax $\pi=0.1$ and for the fitted parallax, $\pi_{\text{fit}}$,
are listed. The positions listed are for the indicated MJDs (epochs
with the best SNR). For S1 the positions were measured relative to
the phase calibrator while the positions of the remaining sources
were measured relative to S1. }

\centering{}%
\begin{tabular}{cccccccccc}
\noalign{\vskip-0.3cm}
\hline\hline &  &  &  &  &  &  &  &  & \tabularnewline
\noalign{\vskip-0.2cm}
\multirow{3}{*}{} &  &  &  &  & $d\,$\textsuperscript{a} &  & $\mu_{\alpha}^{\pi=0.1}$  & $\mu_{\delta}^{\pi=0.1}$ & $\chi_{\text{{red}}}^{2,\,\pi=0.1}$\tabularnewline[\doublerulesep]
\noalign{\vskip\doublerulesep}
 &  &  & Epoch &  & {[}arcsec{]} & $\pi_{\text{fit}}$  & $\mu_{\alpha}^{\pi_{\text{fit}}}$ & $\mu_{\delta}^{\pi_{\text{fit}}}$ & $\chi_{\text{{red}}}^{2,\,\pi_{\text{fit}}}$\tabularnewline[\doublerulesep]
\noalign{\vskip\doublerulesep}
 & Ra (J2000) & Dec (J2000) & (MJD) & SNR & {[}pc{]} & {[}mas{]} & {[}mas$\,$yr$^{-1}${]} & {[}mas$\,$yr$^{-1}${]} & \tabularnewline[\doublerulesep]
\hline 
\noalign{\vskip0.2cm}
\multirow{2}{*}{M15A} & \multirow{2}{*}{21:29:58.246512$\,$(4)} & \multirow{2}{*}{12:10:01.2339$\,$(1)} & \multirow{2}{*}{55146} & \multirow{2}{*}{24} & 1.2$\,$(3) & \multirow{2}{*}{$-$0.02$\,$(10)} & $\mbox{\ensuremath{-}0.54\,(14)}$ & $\mbox{\ensuremath{-}4.33\,(25)}$ & $\mbox{1.4}$\tabularnewline
 &  &  &  &  & 0.06$\,$(1) &  & $\mbox{\ensuremath{-}0.56\,(14)}$ & $\mbox{\ensuremath{-}4.34\,(25)}$ & $\mbox{1.3}$\tabularnewline
\noalign{\vskip0.3cm}
\multirow{2}{*}{M15C} & \multirow{2}{*}{21:30:01.203493$\,$(7)} & \multirow{2}{*}{12:10:38.1592$\,$(2)} & \multirow{2}{*}{55146} & \multirow{2}{*}{15} & 56.0$\,$(3) & \multirow{2}{*}{0.22$\,$(17)} & $\mbox{\ensuremath{-}0.75\,(9)}$ & $\mbox{\ensuremath{-}3.52\,(13)}$ & $\mbox{0.7}$\tabularnewline
 &  &  &  &  & 2.79$\,$(1) &  & $\mbox{\ensuremath{-}0.76\,(10)}$ & $\mbox{\ensuremath{-}3.53\,(15)}$ & $\mbox{0.9}$\tabularnewline
\noalign{\vskip0.3cm}
\multirow{2}{*}{AC211} & \multirow{2}{*}{21:29:58.312403$\,$(4)} & \multirow{2}{*}{12:10:02.6740$\,$(2)} & \multirow{2}{*}{55871} & \multirow{2}{*}{24} & 1.5$\,$(3) & \multirow{2}{*}{0.17$\,$(7)} & $\mbox{\ensuremath{-}0.46\,(8)}$ & $\mbox{\ensuremath{-}4.31\,(20)}$ & $\mbox{3.4}$\tabularnewline
 &  &  &  &  & 0.07$\,$(1) &  & $\mbox{\ensuremath{-}0.49\,(8)}$ & $\mbox{\ensuremath{-}4.32\,(20)}$ & $\mbox{3.4}$\tabularnewline
\noalign{\vskip0.3cm}
\multirow{2}{*}{S1} & \multirow{2}{*}{21:29:51.9034555$\,$(4)} & \multirow{2}{*}{12:10:17.13240$\,$(1)} & \multirow{2}{*}{55352} & \multirow{2}{*}{287} & 95.5$\,$(3) & \multirow{2}{*}{0.04$\,$(25)} & $\mbox{\ensuremath{+}0.06\,(28)}$ & $\mbox{\ensuremath{+}0.53\,(59)}$ & $\mbox{2749}$\tabularnewline
 &  &  &  &  & 4.77$\,$(1) &  & $\mbox{\ensuremath{+}0.05\,(30)}$ & $\mbox{\ensuremath{+}0.53\,(62)}$ & $\text{3034}$\tabularnewline
\noalign{\vskip0.3cm}
\multirow{2}{*}{S2} & \multirow{2}{*}{21:30:02.085700$\,$(8)} & \multirow{2}{*}{12:09:04.2203$\,$(2)} & \multirow{2}{*}{55871} & \multirow{2}{*}{17} & 79.2$\,$(3) & \multirow{2}{*}{0.45$\,$(8)} & $\mbox{\ensuremath{-}0.07\,(13)}$ & $\mbox{\ensuremath{-}1.26\,(29)}$ & $\mbox{2.9}$\tabularnewline
 &  &  &  &  & 3.95$\,$(1) &  & $\mbox{\ensuremath{-}0.05\,(8)}$ & $\mbox{\ensuremath{-}1.27\,(17)}$ & $\mbox{1.0}$\tabularnewline
\hline\hline &  &  &  &  &  &  &  &  & \tabularnewline
\noalign{\vskip-0.3cm}
\multicolumn{10}{c}{{\tiny \textsuperscript{{\tiny a}}Distance (in arcsec and pc) from
the assumed cluster center at RA = 21\tsup{h}29\tsup{s}58\fs330$\pm$0\fs013, Dec = 12\degr10\arcmin01\farcs2$\pm$0\farcs2
\citep{goldsbury10}}}\tabularnewline
\end{tabular}
\end{table*}

\begin{figure*}[t]
\begin{centering}
\subfloat{\raggedright{}\includegraphics[width=0.33\textwidth]{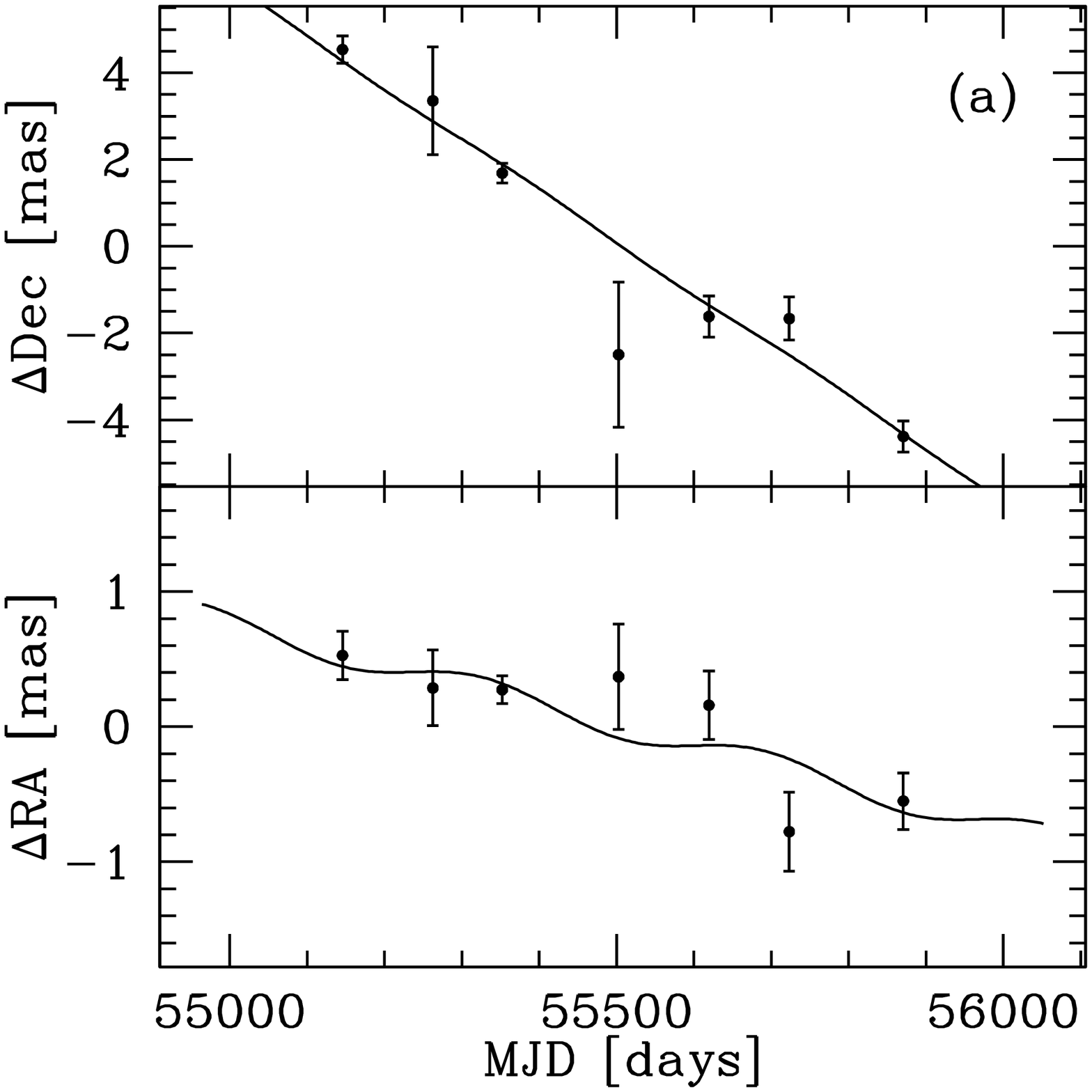}}\subfloat{\centering{}\includegraphics[width=0.33\textwidth]{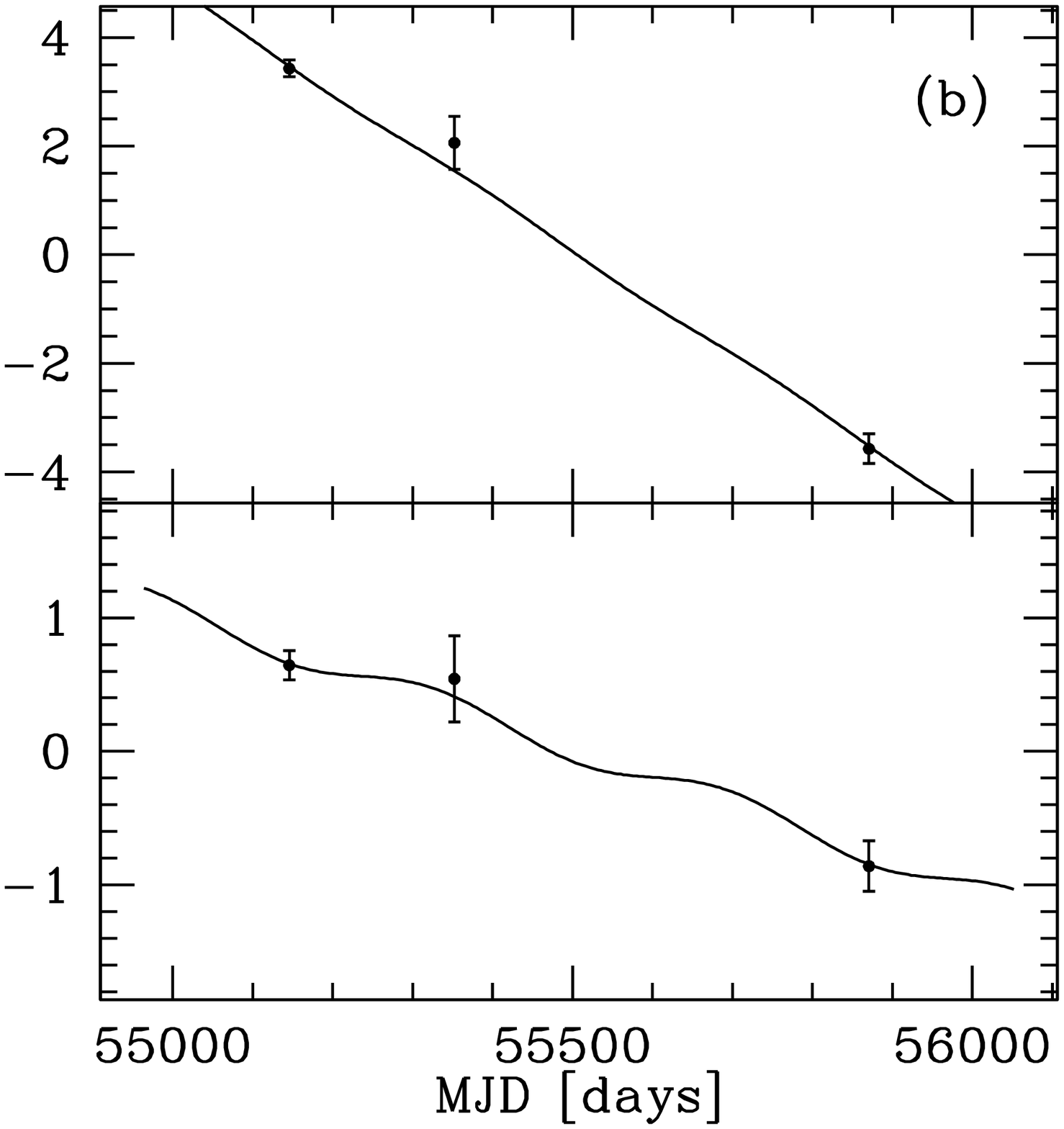}}\subfloat{\raggedleft{}\includegraphics[width=0.33\textwidth]{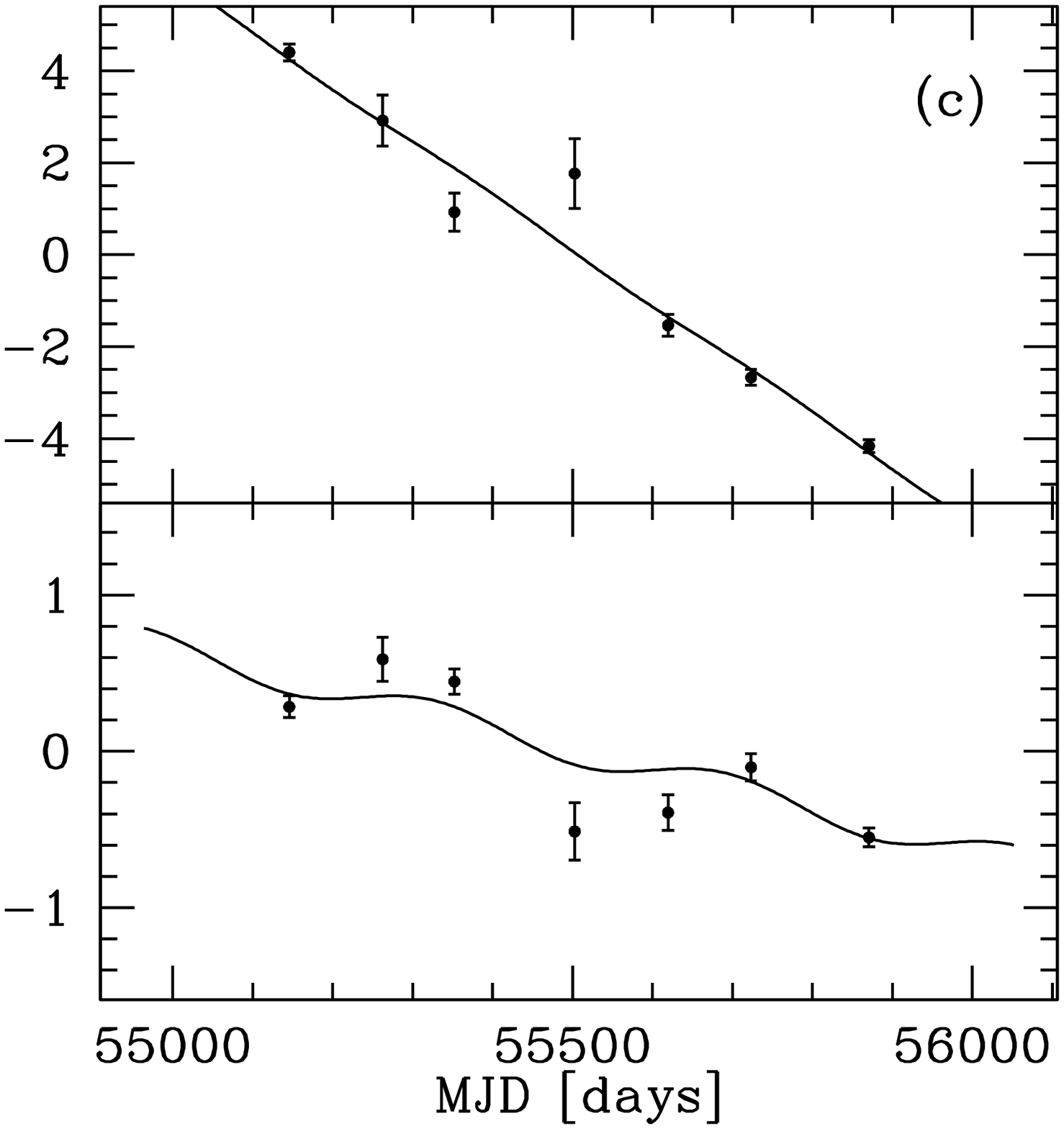}}
\par\end{centering}

\caption{\label{fig:15A,B,AC211-Astrometric-fits}Relative changes in positions
relative to the in-beam calibrator S1 in Dec (top panels) and RA (bottom
panels) for (a) pulsar M15A, (b) pulsar M15C, (c) the LMXB AC211.
The solid line is a fit to position and proper motion keeping the
parallax fixed at $\pi=0.1.$}
\end{figure*}

Once all calibration tables are applied the data of each correlation
center is shifted to nine positions throughout the FOV of each correlation
center using the AIPS task UVFIX. The separation in RA and DEC from
the coordinates of each correlation center are 32.5 arcsec for the
horizontal and vertical positions while it is 22.5 arcsec for the
diagonal positions (Figure \ref{fig:Imaging-strategy}). This data
is averaged to 64 channels per IF and 2 sec integration time (FOV$\sim18$
arcsec for <10\% smearing) before imaging with natural weights. The
images have a dimension of $8192\times8192$ pixels each at a resolution
of 1 mas/pixel. To image one of the nine shifted and averaged data
sets we produce 25 images whose central coordinates are separated
by 8.1 arcsec in right ascension and declination. In total we required
25 images$\times$9 shifts$\times$6 correlation centers$=$1350 images
in each of seven epochs to image the entire primary beam. For source
detection we also compute noise maps for each image with the AIPS
task RMSD. The entire imaging process described above was done in
a parallelized script written in ParselTongue \citep{kettenis06}.

\section{Results\label{sec:Results}}

\begin{figure}
\begin{centering}
\includegraphics[width=0.75\columnwidth]{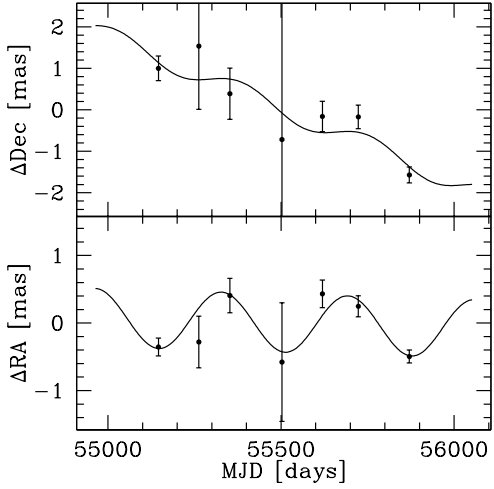}
\par\end{centering}

\caption{\label{fig:S2-Astrometric-fit}The same as Figure \ref{fig:15A,B,AC211-Astrometric-fits}
but for S2. Here we kept the parallax as free parameter.}
\end{figure}

\subsection{Detected sources}

Based on the CLEANed image and its corresponding noise map the AIPS
task SAD was used to search for sources down to an SNR $>4$. In the
following, we cross-correlate all potential candidates in between
epochs allowing for a positional offset $\leq$25$\,$mas. Only potential
candidates found in at least three epochs were followed up for further
analysis. All positions were extracted from a Gaussian fit to the
brightness distribution of each source in the image plane using the
AIPS task IMFIT. Figure \ref{fig:Distribution-of-sources} displays
the distribution of the five detected sources: The two pulsars M15A
(PSR B2127+11A) and M15C (PSR B2127+11C), the LMXB AC211 (4U 2129+12),
and the two unclassified sources S1 and S2 \citep{knapp96}. Except
for M15C we observe all objects in each of the seven epochs. The double
neutron star system M15C was only observable in epochs 1, 3, and 7
for reasons discussed in section \ref{sec:Variability-of-sources}.
We fit the measured positions of all detected sources for parallax
and proper motion. The results for each object are summarized in Table
\ref{tab:Details-of-the-astrometric-fits}.

The two pulsars and the LMXB are known to be members of M15 while
the origins of S1 and S2 are a priori unclear. Based on the well known
distance of M15 ($d=10.3\pm0.4\,\text{{kpc}}$, \citealp{vandenbosch06})
we once fit for both the parallax and the proper motion and we also
fit the data only for the proper motion keeping the parallax fixed
at the expected value of $\pi=0.1\,\text{mas}$. In case of the three
sources known to be cluster members (Figure \ref{fig:15A,B,AC211-Astrometric-fits})
the fit quality achieved is very similar in both approaches. Moreover,
the fitted parallax is in agreement with the expected one for the
given distance within the uncertainties. The positional uncertainties
are estimated by the formal error, beam size/(2{*}SNR), which is appropriate
given the close proximity to the in-beam calibrator S1, relative to
which the positions were measured. The proper motion results do not
change significantly regardless of whether or not we fit for the parallax.
Overall, the proper motions we measure for M15A and M15C agree with
those obtained from pulsar timing by \citet{jacoby06}. 

We applied the same two fitting strategies in the case of S2 where
the fit to the measured positions of the source (Figure \ref{fig:S2-Astrometric-fit})
improves significantly when treating the parallax as free parameter
(see Section \ref{sub:The-Galactic-origin-of-S2}). 

\begin{figure}[t]
\centering{}\includegraphics[width=0.75\columnwidth]{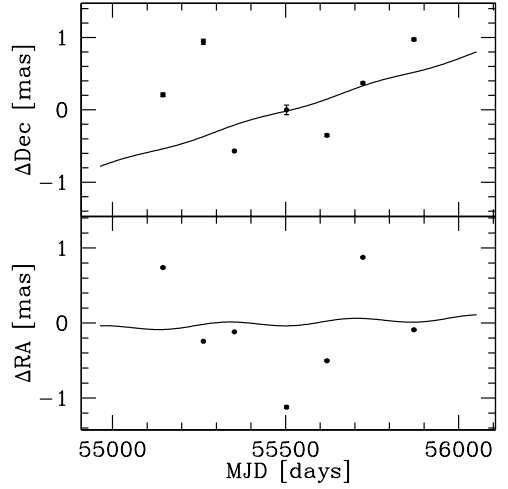}\caption{\label{fig:Relative-positions-of-S1}The same as Figure \ref{fig:15A,B,AC211-Astrometric-fits}
but for the in-beam calibrator S1. Here, the positions are measured
relative to phase calibrator and the parallax is kept as free parameter
in the fit. The error bars are estimated by the formal errors and
are smaller than the data points. See text for details. }
\end{figure}

The astrometric fits for S1 (Figure \ref{fig:Relative-positions-of-S1})
are of very poor quality regardless of fitting strategy. This is indicative
of unmodeled systematic errors, especially since the high flux density
of the source results in very small formal errors ($\sim10\,\mu$as
in RA and $\sim40\,\mu$as in Dec). Given the relatively large angular
separation ($\sim3.\!\!^{\circ}17$) between S1 and the phase calibrator
relative to which the position of S1 is measured, the real uncertainties
may contain a correspondingly large contribution from unmodelled differential
ionospheric phase perturbations along the lines of sight to the two
sources \citep[e.g. ][]{chatterjee09}. To reflect these the formal
errors stated above should be a factor of about $50$ higher. This
would also be in agreement with the trend in the correlation of position
uncertainties and calibrator throw as predicted by Figure 3 in \citet{chatterjee04}.

The implications of the fitting results for S1 and S2 are discussed
in Section \ref{sec:Discussion}.

\subsection{Undetected sources}

Of the eight known pulsars in M15 we do not detect six. Based on the
published flux densities $S_{\nu}$ at $\nu=400\,$MHz (Table \ref{tab:Pulsar-flux-densities})
and our best sensitivity limits (rms$\,\sim3.5\,\mu$Jy, Table \ref{tab:obs-details})
at $\nu=1.6\,$GHz we can, however, derive a lower limit for the spectral
index $\alpha$ of these sources ($S_{\nu}\propto\nu^{-\alpha}$).
Taking the $4\sigma$ upper flux density limit of $S_{1600}\leq14\,\mu$Jy/beam
we find $\alpha\geq1.4$ for all of the undetected pulsars. In the
case of the detected pulsars B2127+11A and B2127+11C our results are
in agreement with the canonical value $\alpha=1.8\pm0.2$ \citep{maron00}
and also with the latest analysis of the pulsar spectral index distribution
suggesting $\alpha=1.4\pm0.96$ \citep{bates2013}. The unusually
high value $\alpha\geq3.0$ for B2127+11B lies even outside the broader
distribution from \citet{bates2013}. Either the intrinsic spectral
index of this pulsar is particularly steep or the cataloged flux density
$\nu_{400}$ was affected by scintillation. Given the close proximity
($\approx5.7''$) of B2127+11B to B2127+11A and the fact that we do
not see any evidence for scintillation for B2127+11A we exclude that
our observations were affected by scintillation.

\begin{table}
\caption{\label{tab:Pulsar-flux-densities}Pulsar flux densities and inferred
spectral index.}

\centering{}%
\begin{tabular}{lccc}
\noalign{\vskip-0.3cm}
\hline\hline &  &  & \tabularnewline
 & $S_{400}$ & $S_{1600}$ & \tabularnewline
 & $[\mu\text{Jy}]$ & $[\mu\text{Jy}]$ & $\alpha$\tabularnewline
\hline 
\noalign{\vskip\doublerulesep}
B2127+11A & 1700\textsuperscript{a} & 100 & 2.0\tabularnewline
B2127+11B & 1000\textsuperscript{b} & $<$14 & $>$3.0\tabularnewline
B2127+11C & 600\textsuperscript{b} & 50 & 1.8\tabularnewline
B2127+11D & 300\textsuperscript{b} & $<$14 & $>$2.2\tabularnewline
B2127+11E & 200\textsuperscript{b} & $<$14 & $>$1.9\tabularnewline
B2127+11F & 100\textsuperscript{b} & $<$14 & $>$1.4\tabularnewline
B2127+11G & 100\textsuperscript{b} & $<$14 & $>$1.4\tabularnewline
B2127+11H & 200\textsuperscript{b} & $<$14 & $>$1.9\tabularnewline
\hline\hline &  &  & \tabularnewline
\noalign{\vskip-0.3cm}
\multicolumn{4}{l}{{\tiny \textsuperscript{{\tiny a}} \citet{wolszczan89} ~~~\textsubscript{}\textsuperscript{{\tiny b}}
\citet{anderson1993}}}\tabularnewline
\end{tabular}
\end{table}

\section{Discussion\label{sec:Discussion}}

\subsection{Implications of proper motion measurements on peculiar velocities
of cluster members\label{sub:Implications-of-proper-motion-on-peculiar-velocity}}

\begin{center}
\begin{table*}
\caption{\label{tab:Global-motion-of-m15}Global motion of M15 and implied
peculiar transverse velocities of the detected cluster members. Case
(i), (ii), and (iii) refer to the average of (i) all three cluster
members; (ii) only M15C; (iii) only M15A and AC211 as proxy for the
global cluster motion.}

\centering{}%
\begin{tabular}{lccccc}
\noalign{\vskip-0.3cm}
\hline\hline &  &  &  &  & \tabularnewline
\noalign{\vskip-0.2cm}
 & $\mu_{\alpha}$ & $\mu_{\delta}$ & M15A  & AC211 & M15C\tabularnewline
{} & {[}mas$\,$yr$^{-1}${]} & {[}mas$\,$yr$^{-1}${]} & {[}km$\,$s$^{-1}${]} & {[}km$\,$s$^{-1}${]} & {[}km$\,$s$^{-1}${]}\tabularnewline
\hline 
\noalign{\vskip\doublerulesep}
\citet{cudworth93} & $-$0.3 $\pm$ 1.0 & $-$4.2 $\pm$ 1.0 & 12 & 9 & 38\tabularnewline
\citet{geffert93} & $-$1.0 $\pm$ 1.4 & $-$10.2 $\pm$ 1.4 & 279 & 280 & 316\tabularnewline
\citet{scholz96} & $-$0.1 $\pm$ 0.4 & $+$0.2 $\pm$ 0.3 & 215 & 214 & 178\tabularnewline
\citet{odenkirchen97} & $-$2.5 $\pm$ 1.5 & $-$8.3 $\pm$ 1.5 & 209 & 212 & 241\tabularnewline
\citet{jacoby06} & $-$1.0 $\pm$ 0.4 & $-$3.6 $\pm$ 0.8 & 40 & 42 & 12\tabularnewline
This work case (i) & $-$0.58 $\pm$ 0.18  & $-$4.05 $\pm$ 0.34 & 13 & 13 & 26\tabularnewline
This work case (ii) & $-$0.75$\pm$0.09 & $-$3.52$\pm$0.13 & 39 & 39 & 0\tabularnewline
This work case (iii) & $-$0.50$\pm$0.16 & $-$4.32$\pm$0.32 & 1 & 1 & 39\tabularnewline
\hline\hline &  &  &  &  & \tabularnewline
\end{tabular}
\end{table*}

\par\end{center}

From the measured proper motions of the sources known to be cluster
members (M15A, M15C, AC211) we can estimate the transverse velocities
of these objects relative to the cluster. Here, we estimate the global
motion of the cluster in three ways: (i) we take the average in $\mu_{\alpha}^{\pi=0.1}$
and $\mu_{\delta}^{\pi=0.1}$ of all three sources; (ii) we take the
proper motion of M15C because of its large angular distance from the
cluster core expecting its motion to be least affected by the gravitational
potential of the central mass concentration; (iii) we take the average
of M15A and AC211 because of their proximity to the cluster core assuming
their peculiar velocity has a radial component only. The results are
summarized in Table \ref{tab:Global-motion-of-m15}.

Case (i) yields the most realistic and most conservative estimate
for the global motion of the cluster. The inferred peculiar velocity
of all three cluster members is well below the escape velocity of
M15 ($v_{\text{esc}}\approx50\,$km$\,$s$^{-1}$ at the position
of M15A and AC211 and $v_{\text{esc}}\approx30\,$km$\,$s$^{-1}$
at the position of M15C, \citealp{phinney93}). We note however, that
the implied peculiar velocity of M15A and AC211 is lower compared
to the one of M15C. Given the close proximity of M15A and AC211 to
the cluster core we would expect the opposite. Since the given global
motion is based on small number statistics this might only be accidental:
The peculiar velocity of M15A and AC211 seem to be very similar and,
therefore, shift the average cluster motion in their favor. Alternatively,
the higher peculiar velocity of M15C might only be a projection effect.
The period derivative of M15A is negative indicating that the pulsar
is being accelerated in the direction of the line of sight \citep{wolszczan89}.
The same could be true for AC211 resulting in a larger 3d-velocity
of M15A and AC211 than that of M15C.

Cases (ii) and (iii) yield the limiting but opposite possible transverse
velocities for all three cluster members. Case (ii) implies a transverse
velocity $v_{\mathrm{\mathrm{trans}}}\sim39\,$km$\,$s$^{-1}$ for
both M15A and AC211 while M15C would only have a radial velocity component.
Taking the uncertainties of the proper motion into account, M15A and
AC211 have a relative transverse velocity of at most $v_{\mathrm{trans}}^{\mathrm{rel}}=27\,$km$\,$s$^{-1}$.
Hence, the highest possible transverse velocity of any of the two
sources $v_{\mathrm{trans}}^{\mathrm{max}}=66\,$km$\,$s$^{-1}$,
significantly above $v_{\text{esc}}$. 

Case (iii) implies hardly any peculiar transverse motion of M15A and
AC211 within the cluster. Given their projected distance from the
core ($0.06\,$pc and $0.07\,$pc for M15A and AC211, respectively)
and assuming a central mass of $3400\,$M$_{\odot}$ \citep{vandenbosch06}
the negligible transverse velocity implies a radial velocity $v_{r}\sim15\,$km$\,$s$^{-1}$
if both sources are on a stable orbit around the cluster core. It
is, however, very unlikely that both sources are on an orbit coinciding
directly with the line of sight to M15. We conclude that case (i)
is the best estimate for the global motion of M15. 

Table \ref{tab:Global-motion-of-m15} lists the previously published
proper motion determinations for the globular cluster M15 as a whole.
All were based on stellar proper-motion measurements, except that
of \citet{jacoby06}, which was based on pulsar timing. The values
of \citet{cudworth93} and of \citet{jacoby06} agree with our results
within the uncertainties. The implied peculiar transverse motions
of M15A, M15C, and AC211 are also below $v_{\text{esc}}$. In combination
with our measurements, the remaining three published results would
indicate velocities relative to the cluster of about $250\,$km$\,$s$^{-1}$
for all three cluster members discussed here. Given the cluster escape
velocity we can exclude these results based on the assumption that
all three compact objects are bound to M15.

\subsection{Undetected pulsars in context of the pulsar luminosity function\label{sub:discussion-luminosity-function}}

The complementary cumulative frequency distribution function (CCFDF)
of pulsar luminosities, $N(\geq L_{\nu})=N_{0}L_{\nu}^{\beta}$, is
a measure for the expected number of pulsars $N$ brighter than the
pseudo luminosity $L_{\nu}=S_{\nu}d^{2}$ (\citealt[and references therein]{bagchi13}).
The\textcolor{red}{{} }parameter $N_{0}$ denotes the number of observable
pulsars above the limiting pseudo luminosity $L_{\nu,\text{min}}=1\,$mJy$\,$kpc$^{2}$
and $\beta$ describes the power law dependence. Here, we adopt the
latest results for the pulsars in M15 of $(N_{0},\,\beta)=(8{}_{-2}^{+3},\,-0.83\pm0.34)$
for $\nu=1400\,$MHz from \citet{bagchi11}. Furthermore, we convert
our upper flux density limit $S_{1600}=14\,\mu$Jy to $L_{1400}=1.9\,$mJy$\,$kpc\textsuperscript{2}
(assuming $\alpha=1.8$ and $d=10.3\,$kpc). Accordingly, we estimate
the number of pulsars we should have observed to $N(\geq L_{1400})=5_{-2}^{+3}$
of which we only detect two. In comparison, \citet{hessels07} report
a limiting pseudo luminosity $L_{1400}=2.1\,$mJy$\,$kpc$^{2}$ which
allows them to detect five pulsars in the GC, well consistent with
the CCFDF predicting $N(\geq L_{1400})=4_{-2}^{+3}$. Despite our
observations being more sensitive, the difference in the number of
detected pulsars can be explained by the fact that \citet{hessels07}
performed pulse searches while our data is pure continuum.

\subsection{Extragalactic origin of S1\label{sub:Extragalactic-origin-of-S1}}

A priori, it was unclear whether the source S1 is a member of M15,
a Galactic back-/foreground source or whether it is of extragalactic
origin. After initial imaging and self-calibration of S1 we perform
the first astrometric analysis. Figure\textcolor{black}{{} \ref{fig:Relative-positions-of-S1}
shows the position of the source relative to the phase calibrator
in each epoch. The measured positions scatter about a central position
with an offset of }$\sim1$ mas (= 1 pixel in the image plane). A
fit to parallax, $\pi$, and proper motion, ($\mu_{\alpha},\,\mu_{\delta}$),
yields $\pi=0.04\pm0.25\,\text{mas}$, $(\mu_{\alpha},\mu_{\delta})=(0.05\pm0.30,\,0.53\pm0.62)\,$mas$\,$yr$^{-1}$.
With parallax and proper motion measurements consistent with zero,
an extragalactic origin for S1 is the most natural conclusion. Within
the uncertainty of $\pi$, however, the hypothesis that S1 is related
to the cluster cannot be excluded. Therefore, we check the alternate
hypothesis that S1 is a cluster member by considering its proper motion
relative to that of the GC. Using case (i) from Table \ref{tab:Global-motion-of-m15}
for the GC and adding the proper-motion uncertainties for the two
objects in quadrature, this relative proper motion evaluates to $(\mu_{\alpha}^{\text{rel}},\mu_{\delta}^{\text{rel}})=(0.63\pm0.35,\,4.58\pm0.70)\,$mas$\,$yr$^{-1}$,
which in turn translates to a transverse velocity of S1 relative to
the GC of $220_{-32}^{+30}\,$km$\,$s$^{-1}$. This value is much
larger than t\textcolor{black}{he escape velocity at its position}
($24.1\,\mathrm{km\,}\mathrm{s}^{-1}<v_{\text{esc}}<29.3\,$km$\,$s$^{-1}$,
\citealp{phinney93}) and rules out that S1 and M15 are related.

\subsection{Galactic origin of S2\label{sub:The-Galactic-origin-of-S2}}

In Table \ref{tab:Details-of-the-astrometric-fits} we list the results
of both fitting strategies with and without parallax. If the parallax
is kept fixed at $\pi=0.1\,\text{mas}$, suggesting S2 to be a member
of M15, we measure a proper motion $(\mu_{\alpha,}\mu_{\delta})=(-0.07\pm0.13,-1.26\pm0.29)\,$mas$\,$yr$^{-1}$.
These values are significantly different from the ones we measure
for the cluster members. Adopting our case (i) scenario for the global
motion of M15 (Table \ref{tab:Global-motion-of-m15}), the measured
proper motion would translate to a peculiar transverse velocity $v_{t}^{\text{S2,pec}}=142_{-3}^{+4}\,$km$\,$s$^{-1}$
at PA$\,=10$\degr, opposite to the direction of motion of M15 (PA$\,=188$\degr).
At the angular distance of S2 to the cluster core (Figure \ref{fig:Distribution-of-sources},
Table \ref{tab:Details-of-the-astrometric-fits}) this velocity would
indicate that S2 is not bound to M15 ($24.1\,\mathrm{km\,}\mathrm{s}^{-1}<v_{\text{esc}}<29.3\,$km$\,$s$^{-1}$,
\citealp{phinney93}). Hence, it seems unlikely that S2 is related
to the cluster.

If we take the parallax as free parameter the results for the proper
motion remain the same within the uncertainties but the fit quality
increases by a factor of 3. The change in $\chi_{\text{red}}^{2}$
can be quantified in an F-test \citep[e.g.][]{bevington69,stuart94}.
Given the observed $F$ -- the ratio between the reduced chi-squares
resulting from the fits in which parallax was (i) fixed at 0.1 mas
and (ii) treated as a fittable parameter, which here is 2.9 -- and
the different numbers of degrees of freedom, $\nu_{1}=10$ (fixing
$\pi=0.1$) and $\nu_{2}=9$ (fitting for $\pi$) we can compute the
probability $P_{F}$ of random data producing such an improvement
in the value of reduced chi-square in the latter fit:

\[
P_{F}(F;\,\nu_{1},\,\nu_{2})=\int_{F}^{\infty}P_{f}(f;\,\nu_{1},\,\nu_{2})\,\mathrm{df},
\]

with
\[
P_{f}(f;\,\nu_{1},\,\nu_{2})=\frac{\Gamma[\frac{\nu_{1}+\nu_{2}}{2}]}{\Gamma[\frac{\nu_{1}}{2}]\Gamma[\frac{\nu_{2}}{2}]}\cdot\left(\frac{\nu_{1}}{\nu_{2}}\right)^{\frac{\nu_{1}}{2}}\cdot\frac{f^{\frac{\nu_{1}-2}{2}}}{(1+f\frac{\nu_{1}}{\nu_{2}})^{\frac{\nu_{1}+\nu_{2}}{2}}}\,.
\]

Evaluating $P_{F}(2.9;\,10,\,9)$ yields 0.062. Hence, we can reject
the hypothesis of random data producing a ratio $F\geq2.9$ at the
94\% confidence level. Therefore, we conclude that the measured parallax
$\pi=0.45\pm0.08\,\text{mas}$ is real. This value indicates that
S2 is of Galactic origin and, furthermore, that it is a foreground
source at a distance $d=2.2_{-0.3}^{+0.5}\,$kpc. After correcting
for solar motion (with the relevant solar parameters taken from \citealp{schoenrich12})
and also for differential galactic rotation (with the Oort constants
A and B as published by \citealp{feast97}) our measured proper motion
$(\mu_{\alpha,}\mu_{\delta})=(-0.05\pm0.08,-1.27\pm0.17)\,$mas$\,$yr$^{-1}$
(Figure \ref{fig:S2-Astrometric-fit}) implies a transverse velocity
$v_{t}^{\text{S2}}=26_{-4}^{+5}\,$km$\,$s$^{-1}$ with respect to
the local standard of rest (LSR) at position angle PA$\,=-51$\degr
(PA in equatorial coordinates). We discuss the nature of S2 in the
following section.

\subsection{Variability of sources\label{sec:Variability-of-sources}}

\begin{figure}[b]
\begin{centering}
\includegraphics[width=0.95\columnwidth]{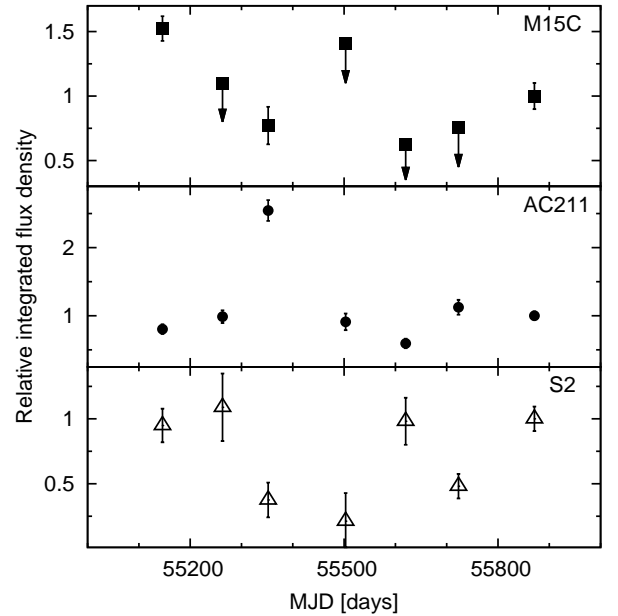}
\par\end{centering}

\caption{\label{fig:Flux-evolution} Relative integrated flux density evolution
of significantly variable sources normalized to the integrated flux
density in epoch 7 ($32\pm3\,\mu$J, $224\pm14\,\mu$J, $214\pm20\mu$J
for M15C, AC211, and S2, respectively). From top to bottom: M15C,
AC211, S2. For the pulsar M15C only $4\sigma$ upper flux density
limits are indicated for epochs 2, 4, 5, and 6. For AC211 in epoch
3 the flux is integrated over both components.}
\end{figure}

\begin{figure}
\begin{centering}
\includegraphics[width=0.9\columnwidth]{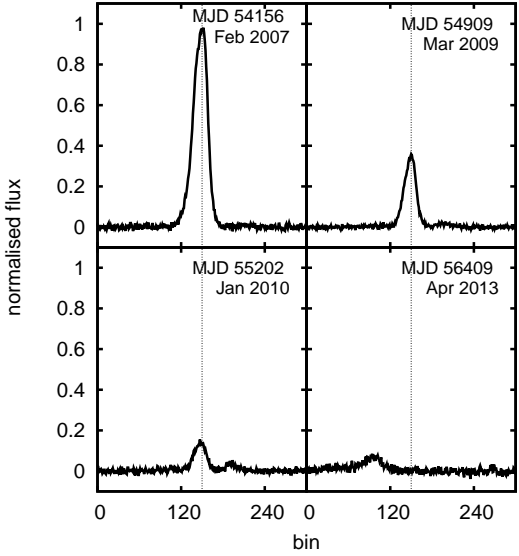}
\par\end{centering}

\caption{\label{fig:Profile-evolution-of-15C}Pulse profile evolution of M15C
as observed at Arecibo. The pulsar slowly faded away over the course
of about three years, probably due to geodetic precession. The horizontal
dotted line indicates the central location of the peak as observed
until January 2010. The shift in the location of the central peak
in the bottom right panel is consistent with a 2.5\% phase shift observed
in the timing observations (only the relevant $\sim10$\% of the pulse
phase are shown here). This shift is indicative of an emission component
different than the one observed up to 2010 that moved into our line
of sight due to geodetic precession (Ridolfi et al. 2014, in preparation).
There were no observations at Arecibo between MJD 55202 and MJD 56409.}
\end{figure}

\textbf{M15C}\\
The proper motion results for the double neutron star system M15C
are based on three observations only (epochs 1, 3, and 7, Figure \ref{fig:15A,B,AC211-Astrometric-fits}(b)).
The non-detection in epochs 2 and 4 is mostly due to the fact that
those epochs were observed without the Arecibo dish which is essential
to obtain the sensitivity required to detect this system. Apart from
this fact, however, the measured flux density of M15C decreased between
epochs 1 and 3 only to fade beyond detection as of epoch 5 (Figure
\ref{fig:Flux-evolution}). In epoch 7 the pulsar was observable again.
To confirm the reappearance of the source we performed single dish
follow-up observations with the Arecibo telescope which had previously
stopped monitoring the system due to its low flux density. Previously,
regular timing observations had also revealed a steady decrease in
peak flux density and also a change in pulse profile (Figure \ref{fig:Profile-evolution-of-15C},
Ridolfi et al. 2014, in preparation). Our follow-up observations (bottom
right panel of Figure \ref{fig:Profile-evolution-of-15C}) confirm
that the emission we detect in epoch 7 is related to M15C. They also
reveal, however, a shift in pulse peak location that is consistent
with a 2.5\% phase shift observed in the timing observations (Ridolfi
et al. 2014, in preparation). A possible explanation for this behavior
is geodetic precession: due to orbit-spin coupling the axis of rotation
of the one neutron star we see as a pulsar precesses, moving the emission
region out of our line of sight. The same effect then moves a different
component of the emission cone into our line of sight, which we detect
at a slightly different phase. \\
\\

\begin{figure}
\begin{centering}
\includegraphics[width=0.9\columnwidth]{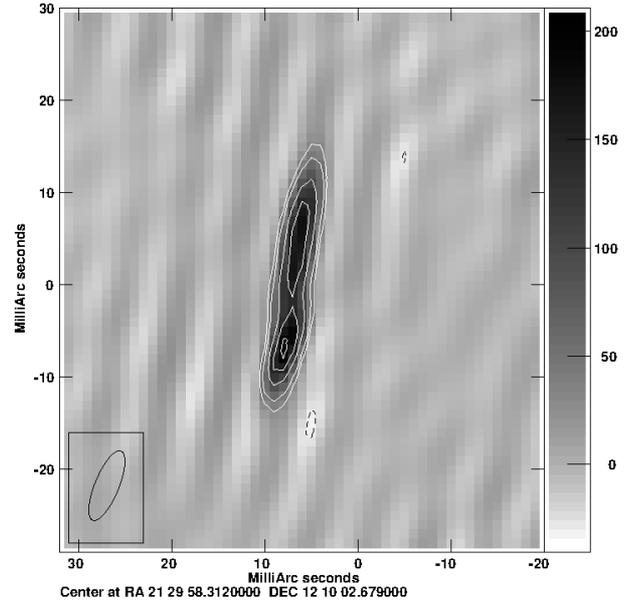}
\par\end{centering}

\caption{\label{fig:AC211-E3-kntr}Clean image of AC211 in epoch 3 (MJD 55352)
exhibiting the double lobed structure. Contours are (-3, 3, 5, 10,
15, 20) times the noise level $rms=9.5\,\mu\text{Jy/beam}$. The beam
size and position angle are shown in the bottom left corner.}
\end{figure}

\textbf{AC211}
\begin{figure}
\includegraphics[width=0.95\columnwidth]{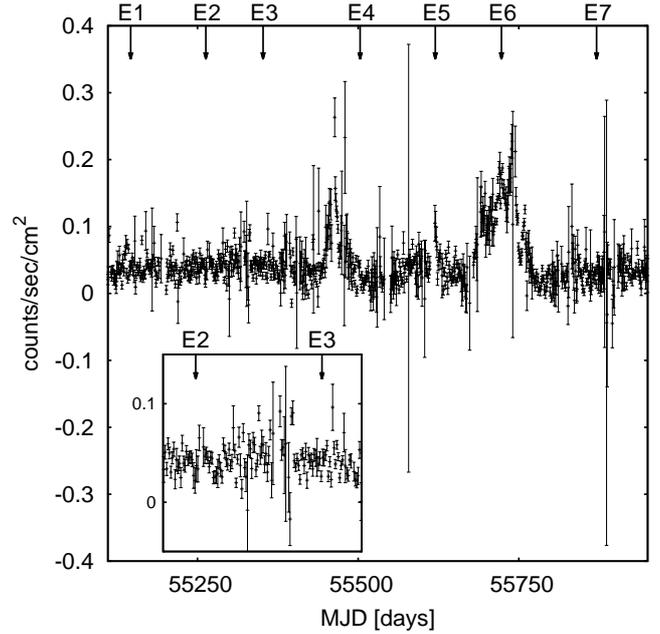}

\textbf{\caption{\label{fig:X-ray-lightcurve-of-M15}X-ray lightcurve (2-20 keV) of
M15 from the MAXI mission covering the time of this campaign. The
inset is a blow-up of the timerange between epochs 2 and 3. Note the
slight increase in count rate about 40 days prior to epoch 3.}
}

\end{figure}

The integrated flux density of AC211 varied slightly about an average
value of $\sim200\,\mu\text{Jy}$. This is true for all observations
but epoch 3. In this observation at MJD 55352 we measure an integrated
flux density $S_{int}=570\pm34\,\mu\text{Jy}$ (Figure \ref{fig:Flux-evolution}).
Moreover, in this epoch the source exhibits a double lobed structure
extending from north to south (Figure \ref{fig:AC211-E3-kntr}). We
exclude the possible explanation of this structure by a phase error
for two reasons: (i) When imaged, the close-by pulsar M15A (located
at about 1.7 arcsec to the south-west of AC211) is an unresolved point
source also in epoch 3. Any phase errors causing the observed structure
in AC211 should result in a similar structure also for M15A. (ii)
We can reproduce the structure independent of the data reduction and
calibration strategy.

Therefore, we conclude that the observed bipolar structure is real
and might have been caused by an outburst that occurred at some time
between epochs 2 and 3. The X-ray data published by the \textit{MAXI}
mission%
\footnote{http://maxi.riken.jp%
} \citep{matsuoka09} reveals X-ray variability of M15 during our observations
that could be caused by the activity of AC211 (Figure \ref{fig:X-ray-lightcurve-of-M15}).

At the distance of M15 the angular separation between the components
corresponds to about 137 AU. The time span from epoch 2 to 3 is 110
days, the one from epoch 3 to 4 is 150 days. Assuming that the outburst
occurred shortly after epoch 2 the transverse velocity $v$ of the
ejected material (ejected from an object in the middle of both components)
would be roughly 0.6 AU per day corresponding to $v\approx1000\,\text{km/s}$.
This velocity is very low compared to measured relativistic jet velocities
of, e.g., the X-ray binaries SS433 ($v\sim0.2$c, \citealp{stirling02},
and references therein) and GRS1915+105 ($v\sim0.98$c, \citealp{fender99},
and references therein). Therefore, our measured velocity could well
be real, especially if the outburst occurred at some time close to
epoch 2. Turning the argument around, the outburst could have occurred
only two days before our observations if we assume an ejection velocity
of $v\sim0.2$c.

In \citet{fender99} it also becomes clear that the radio flux density
of individual components can decrease rapidly on the timescale of
a few months. Hence, the time span of 150 days from epoch 3 to 4 is
long enough for the ejected material to dissipate and become undetectable
again.\\
\\
\textbf{S2}\\
During the course of this campaign the measured flux density of S2
varied by a factor of six between $\sim40$ and $\sim230\,\mu\text{Jy}$
(Figure \ref{fig:Flux-evolution}). \citet{knapp96} report a flux
density of $230\pm40\,\mu\text{Jy}$ at 8.4 GHz and estimate a $5\sigma$
upper limit of $150\,\mu$Jy at $4.9\,$GHz from archival data of
\citet{machin90}. This is indicative of a compact flat spectrum source
variable on the timescale of a few months, most likely an LMXB. To
our knowledge, however, there are no known X-ray sources within $\sim1\,$arcmin
to the coordinates of S2.

\subsection{Constraints on the IMBH mass in M15\label{sub:Constraints-on-the-IMBH}}

In \citet{kirsten2012} we reported a $3\sigma$ upper mass limit
for the putative IMBH in M15 of $M_{\bullet}\leq500\,$M$_{\odot}$.
This work was based on the first five of the observations discussed
here. In epochs 6 and 7 we also do not detect any significant emission
within a radius of $0.6$ arcsec (corresponding to a $3\sigma$ uncertainty)
of the cluster core position at coordinates (J2000) RA = 21\tsup{h}29\tsup{s}58\fs330$\pm$0\fs013, Dec = 12\degr10\arcmin01\farcs2$\pm$0\farcs2
\citep{goldsbury10}. If we concatenate the data of all seven epochs
the resulting noise level improves only slightly compared to concatenating
the first five epochs only (rms$\,\sim3.3\,\mu\text{Jy/beam}$). Based
on the fundamental plane of black hole activity \citep{merloni03,falcke04,kording06}
\[
\text{\text{log}}M_{\bullet}=1.55\,\text{log}L_{R}-0.98\,\text{log}L_{X}-9.95
\]
which relates black hole mass, $M_{\bullet}$; radio luminosity, $L_{R}$;
and X-ray luminosity, $L_{X}$. We therefore reconfirm the upper mass
limit for the putative IMBH of $M_{\bullet}\leq500\,$M$_{\odot}$.
Furthermore, our data excludes variability of any compact object residing
at the center of M15 on the timescale of two months to two years.

\section{Conclusions\label{sec:Conclusions}}

We observed the massive globular cluster M15 in a multi-epoch global
VLBI campaign in seven observations covering a time span of two years.
In our observations we clearly detect five compact radio sources,
namely the pulsar M15A, the double neutron star system M15C, the LMXB
AC211, and two unclassified sources S1 and S2. Except for M15C (which
was only detected in epochs 1, 3, and 7), all sources were detected
in all seven epochs. From our proper motion measurements (Table \ref{tab:Details-of-the-astrometric-fits})
and the variability of M15C, AC211, and S2 we conclude:
\begin{itemize}
\item The projected global proper motion of M15 is $(\mu_{\alpha},\,\mu_{\delta})=(-0.58\pm0.18,\,-4.05\pm0.34)\,$mas$\,$yr$^{-1}$,
\item M15A and AC211 have a maximal transverse peculiar velocity $v_{trans}^{max}=66\,$km$\,$s$^{-1}$
within the cluster,
\item In epoch 3, the morphology of the LMXB AC211 is not point like but
shows a double lobed structure instead. It is quite likely that the
source had an outburst shortly before the observations in epoch 3,
\item M15C has a transverse velocity of at most $39\,$km$\,$s$^{-1}$
moving towards the north in the cluster,
\item The observed 2.5\% phase shift in the pulse profile points to geodetic
precession as a possible explanation for the disappearance and reappearance
of M15C during the observations,
\item S1 is of extragalactic origin, most probably a background quasar,
\item S2 is a Galactic foreground source at a distance $d=2.2_{-0.3}^{+0.5}\,$kpc
moving at a transverse velocity $v_{t}^{\text{S2}}=26_{-4}^{+5}\,$km$\,$s$^{-1}$
with respect to the LSR,
\item The flux density of S2 is variable by a factor of a few on the time
scale of a few months. The spectrum seems to be flat indicative of
a LMXB. There is, however, no known X-ray source within about 1 arcmin
of the radio position of the source.
\end{itemize}
The proper motions measured here will be important for the analysis
of the timing data from M15A and M15C (Ridolfi et al., in preparation).
Our model-independent measurement of the proper motion of the pulsar
M15A will allow a much less ambiguous interpretation of the variation
of the acceleration of this pulsar in the cluster potential. Equally,
in the case of M15C, with timing only the glitch signal will be entangled
with the proper motion signal. Our measurement of the proper motion
will allow an unambiguous study of the rotational behavior of the
pulsar. 

Similar to the first five observations \citep{kirsten2012}, in epochs
6 and 7 we do not detect any significant emission from a putative
IMBH within the central 0.6 arcsec of the core region of M15. Excluding
any variability of a central object on the time scale of two months
to two years, we reconfirm the $3\sigma$ upper limit for the proposed
central IMBH mass of M$_{\bullet}=500\,\text{M}_{\odot}$ .
\begin{acknowledgements}
We appreciate the comments of the anonymous referee that helped us
to improve the manuscript. We would like to thank the JIVE staff for
technical support throughout the observations. Also, we appreciate
the help of Adam Deller who supplied us with the code for the accurate
uv-shifting without which the data of epoch 1 could not have contributed
to this project. Furthermore, we acknowledge the help of André Offringa
helping us in developing a suitable flagging strategy for the AOFlagger.
F.K. acknowledges partial support through the Bonn-Cologne Graduate
School of Physics and Astronomy. The research leading to these results
has received funding from the European Commission Seventh Framework
Programme (FP/2007-2013) under grant agreements No. 227290 (Advanced
Radio Astronomy in Europe) and No. 283393 (RadioNet3). The European
VLBI Network is a joint facility of European, Chinese, South African
and other radio astronomy institutes funded by their national research
councils.
\end{acknowledgements}
\bibliographystyle{aa}
\bibliography{biblio}

\end{document}